\documentstyle[11pt,aaspp4,psfig]{article}  

\begin{document}
 
\title{VLA\footnote{The Very Large Array is a facility of the
National Radio Astronomy Observatory which is operated by Associated
Universities, Inc.\ under cooperative agreement with the National
Science Foundation.}
Observations of Radio Variability of ER Vulpeculae in 1995}

\author{\sc Slavek M. Rucinski\altaffilmark{2}\\
\rm Electronic-mail: {\it rucinski@cfht.hawaii.edu\/}}

\affil{Canada -- France -- Hawaii Telescope Co.\\
P.O.\ Box 1597, Kamuela, HI 96743}

\bigskip
\centerline{\today}

\bigskip
\begin{abstract}
VLA continuous monitoring of ER~Vul during two 11.5 hour sessions
in September 1995 in the radio continuum at 3.6 cm (X-band) showed
a variation pattern similar to that observed in 1990 and 1991, but
at a lower mean flux level. Simultaneous
observations with the EUVE satellite in the 70 -- 140 \AA\ band
did not show any significant flaring variability of the star.
\end{abstract}

\section{INTRODUCTION}
\label{intro}

ER Vulpeculae (HD~200391) is a close, short-period (0.7 day) binary
system consisting of two solar-type stars locked into synchronous
rotation with the rotation rate some 40-times faster than the Sun.
The stars are -- as expected -- very active and show many signatures of
magnetic activity, manifested by strong chromospheric and coronal
emissions as well as rapidly-evolving,
dark photospheric spots. The system was
subject to several analyses of its activity, usually in studies
including other similar, very active, strongly spotted, short-period 
binary systems. One of the most recent photometric analyses specifically 
of this system (which should be referred for previous literature) 
was by Olah et al.\ \markcite{ola94} (1994). An extensive spectroscopic
study of the system was presented by Hill et al.\ \markcite{hil90}
(1990).

Of all signatures of activity in active stars, radio emission and 
its variability are the least explored due to very low
observed fluxes and the implied necessity of using large radio telescopes.
In the case of ER~Vul, about $10^{-8}$ of the
bolometric luminosity is converted
into the radio emission (Rucinski \markcite{ruc92} 1992). Although
this bolometric-to-radio
flux conversion is some $10^5$ times larger than for the solar case
and the system is very active and radio-bright, as expected,
the observed fluxes are low; for the distance of ER~Vul the flux levels
are typically $1 - 10$ mJy. Because of these 
low fluxes, studies of rapid variability require application of
100-meter class radiotelescope, such as the VLA.

ER Vul was observed before with the VLA radiotelescope system in 
1990 and 1991 in 
a variability-monitoring program described in Rucinski \markcite{ruc92} 
(1992 = R92). This reference should be consulted for earlier literature
on the subject and for several details which are omitted
in the current paper. The continuum spectral bands were 20 cm (L-band)
and 6 cm (C-band) in 1990, whereas most of the 1991 observations
were at 3.6 cm (X-band) with only single flux-level checks at 6 cm.
Each time, about 23 hours of monitoring were used, in two
continuous runs. During each run, the star showed a complex variability
pattern, apparently a mixture of short flares typically lasting minutes
and continuous variations with time scales of hours. The latter 
seemed to be partly related to the orbital phases and thus possibly due
to optical-depth effects in a complex magnetosphere surrounding
both stars. The fluxes at 3.6 and 6 cm were very similar in 1991
suggesting a flat radio spectrum. A drop in the overall radio
activity level was noted between 1990 and 1991.

This paper describes 3.6 cm VLA observations of ER~Vul in 1995
conducted in a very similar way as in 1990 and 1991. However,
in contrast to the previous VLA runs, which did not have any
simultaneous support, the 1995 run was scheduled during an
Extreme Ultraviolet Explorer (EUVE) program which lasted one week
(Rucinski \markcite{ruc97} 1998) and provided extreme-ultraviolet
spectra of the star. During these observations, the photometric
Deep Sky Survey channel (70 -- 140 \AA) was activated, providing an
opportunity of seeking correlations between variabilities in thermal
(EUV) and non-thermal (radio) components of the coronal emission.
The results are described in Section~\ref{results} following the
Section~\ref{obser} about the observations.

\section{VLA OBSERVATIONS}
\label{obser}

The 1995 observations reported in this note were done almost
exactly in the same way as in 1991, that is only at 3.6~cm (X-band) 
and during two 11.5 hour continuous runs. The standard VLA
continuum frequencies centered at 
8.415 and 8.465 GHz with bandwidths of 50 MHz
were used. The main differences relative to the 1991 run
were twofold: (1)~because of the VLA and the EUVE satellite scheduling
constraints, the two sessions were not on consecutive
days, but were separated by 3 days; (2)~the planned, 
single, 6~cm (C-band) observations to check the X-band/C-band
relative fluxes were inadvertently omitted precluding verification
that the C- and X-band fluxes were again comparable in 1995, as they
had been in 1991.

\begin{figure}           
\centerline{\psfig{figure=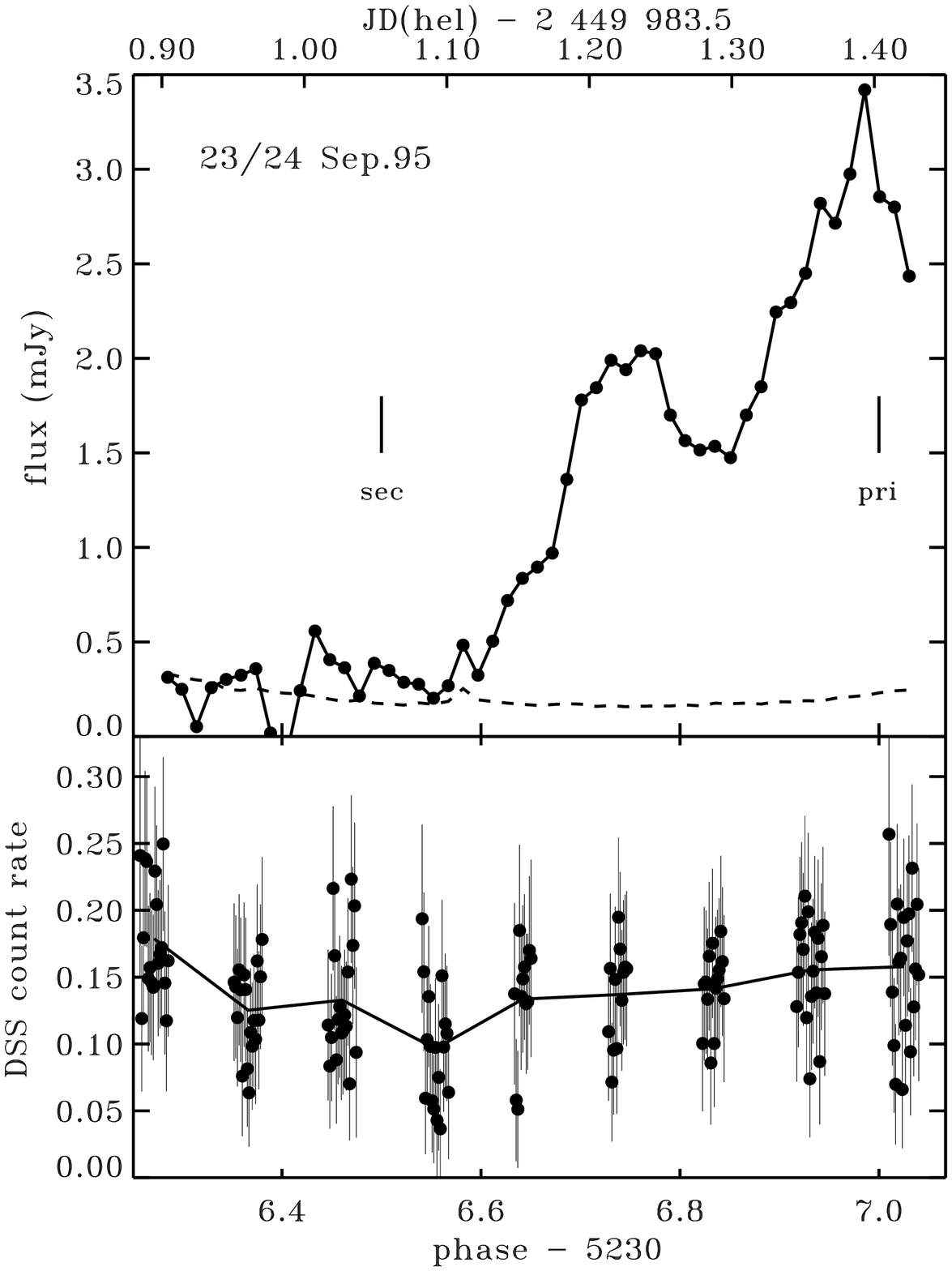,height=4.5in}}
\vskip 0.5in 
\caption{\label{fig1}
Changes of the 3.6 cm continuum flux of ER~Vul are shown in the
upper panel. The estimated errors of observations are plotted
as a broken line. The time axis is in orbital phases counted
from an arbitrary epoch of the primary eclipse, as described in
the text. The upper horizontal axis gives the time in days since
$0^h$~UT on September 23, 1995. The lower panel shows EUVE
observations in the 70 -- 140 \AA\ band, binned in 100 second
intervals and expressed as count rates per second. The error
bars have lengths equal to two standard errors. The line segments
connect the mean count-rate values in groups of data points for
each orbital visibility period of the EUVE satellite.
}
\end{figure}

\begin{figure}           
\centerline{\psfig{figure=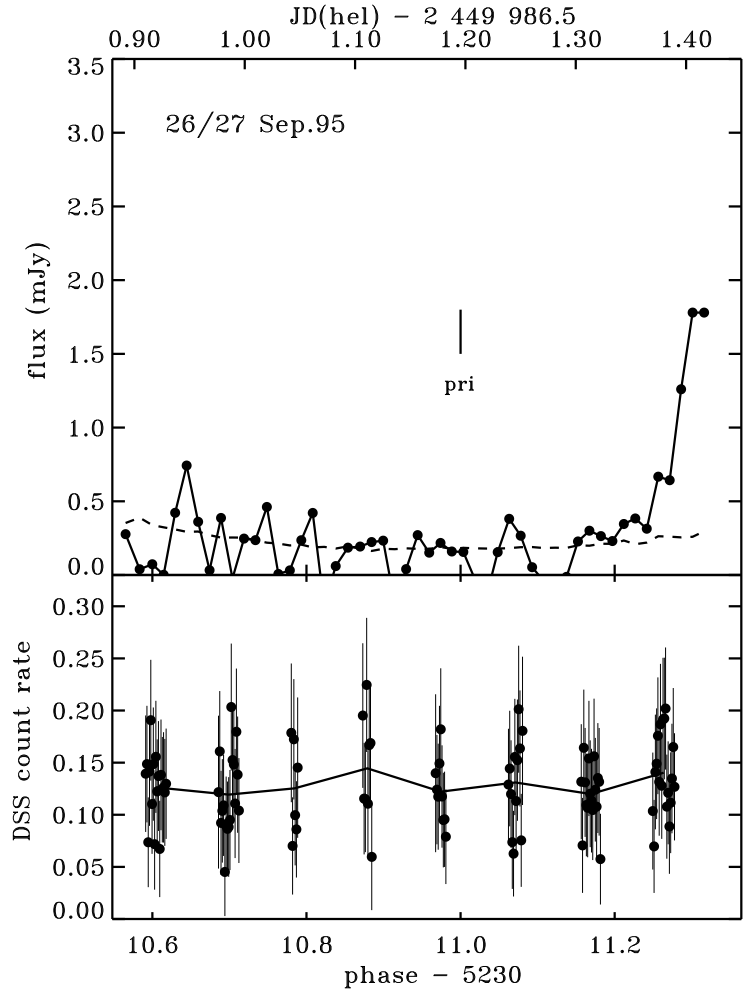,height=4.5in}}
\vskip 0.5in
\caption{\label{fig2}
The same as in Figure~\ref{fig1}, but for September 26, 1995.
}
\end{figure}

The observations were made in equal time intervals of about 15 minutes
with the calibration of the scans against the same phase calibrator
2113+293. Individual scans, about 11 -- 12 minutes long, were analyzed
using the AIPS\footnote{Astronomical Image Processing System (AIPS) is
a data-reduction software packages developed by National Radio Astronomy
Observatory.}. Cleaned radio maps were obtained for all scans; then
radio emission fluxes were extracted from the maps assuming a point source
at the position of ER~Vul. Two different two-dimensional
Gaussian fitting routines JMFIT and IMFIT were used for the flux
extraction. The data in the two 50 MHz spectral bands 
were handled separately to obtain quasi-external
estimates of the flux errors; the reason for this approach was the
fact that the two Gaussian fitting routines are known to give 
unrealistically small flux errors (R92). Estimation of the 
errors in this way led to a conclusion that the the peak-flux
errors provided by the JMFIT were internally consistent, but that they
required an upward scaling by a factor of 4 times for the first run
and 4.5 times for the second run. The fluxes tabulated in
Table~\ref{tab1} are the average values from both 
50 MHz datasets. They are
plotted in the upper panels of Figures~\ref{fig1} and \ref{fig2}
(continuous lines) together with the estimates of the errors (broken
lines). The data were not analyzed for circular 
polarization due to the low flux levels; the previous observations
(R92) did not show any polarization of the signal.

The times given for the scan mid-points 
in Table~\ref{tab1} are the heliocentric
Julian Days based on the International Atomic Time scale (IAT). 
To correlate the stellar variability with the orbital orientation of
ER~Vul, the time shown in Figures~\ref{fig1} and \ref{fig2} was
also expressed in orbital phases counted from the same initial
epoch as for the EUVE observations (Rucinski \markcite{ruc97} 1998).
The ephemeris for the primary (deeper) eclipses was that of Hill
et al.\markcite{hil90} (1990): JD(hel) = 2446328.9837 + 0.698095~E, 
which is apparently still valid, as had been verified one year before
our observations by Zeinali et al.\markcite{zei95} (1995). 
For consistency with the EUVE observations, the epoch E = 5230 was used
as the start of the phase count in Figures~\ref{fig1} and \ref{fig2}.

\section{RESULTS}
\label{results}

The pattern of variability seen during the first run on September 23/24, 
1995 was very similar to that seen in the 1990 and 1991 
observations, with changes taking place in an hourly time-scale,
that in is some 0.1 -- 0.2 of the orbital period.
The flux levels were very low, close to the typical measurement
errors (at or below about 0.3 mJy)
 at the beginning of the run, but then climbed to about
3 mJy at the end of this run which coincided with the primary eclipse
of the binary system. Three days later, on September 26/27, 1997,
practically no emission was observed for long time at the orbital phases
around the primary eclipse; later, 
a slight increase was observed at the very end of this day. Thus,
during this series of observations, practically no relation to 
the orbital phases was seen, contrary to what had been observed in 1990
and 1991.

The DSS data in the 70 -- 140 \AA\ continuum which were obtained
during the radio observations are shown in the lower
panels of Figures~\ref{fig1} and \ref{fig2}. The sampling interval
used for the DSS photon accumulation was 100 seconds. Such
intervals were grouped together by the satellite
orbital-visibility periods.
The short line segments in the figures connect mean values for
such groups in an attempt to accentuate any hourly time-scale 
variability in the DSS count rate. Although small EUV flares were observed
during the whole EUVE run, no significant
variability was  detected during the portion which overlaped
with the radio observations. The DSS data turned out to be consistent 
with a steady count rate of about 0.14 counts/second. 
Apparently, ER~Vul was observed during a time when its magnetic
activity was generally at a low state.

We note that the DSS light curve obtained during the whole duration of the
EUVE run (Rucinski 1998)
showed a mild increase in the quiescent level in the phase
interval between the primary and secondary eclipses, with a few localized,
flare-like increases toward the end of the other half of the orbital
period. The active half of the DSS light curve could be related to the 
highly variable optical light curve in the same phase
interval, as observed by Olah et al. 
\markcite{ola94} (1994) in 1990 -- 1992 and possibly also by 
Zeinali et al.\markcite{zei95} (1995) one year before the 
VLA observations. We see nothing unusual in this phase interval
in the 1995 VLA observations. To the contrary, the radio flux
was the lowest in the phase interval between 
the primary and secondary eclipses. 

\begin{figure}           
\centerline{\psfig{figure=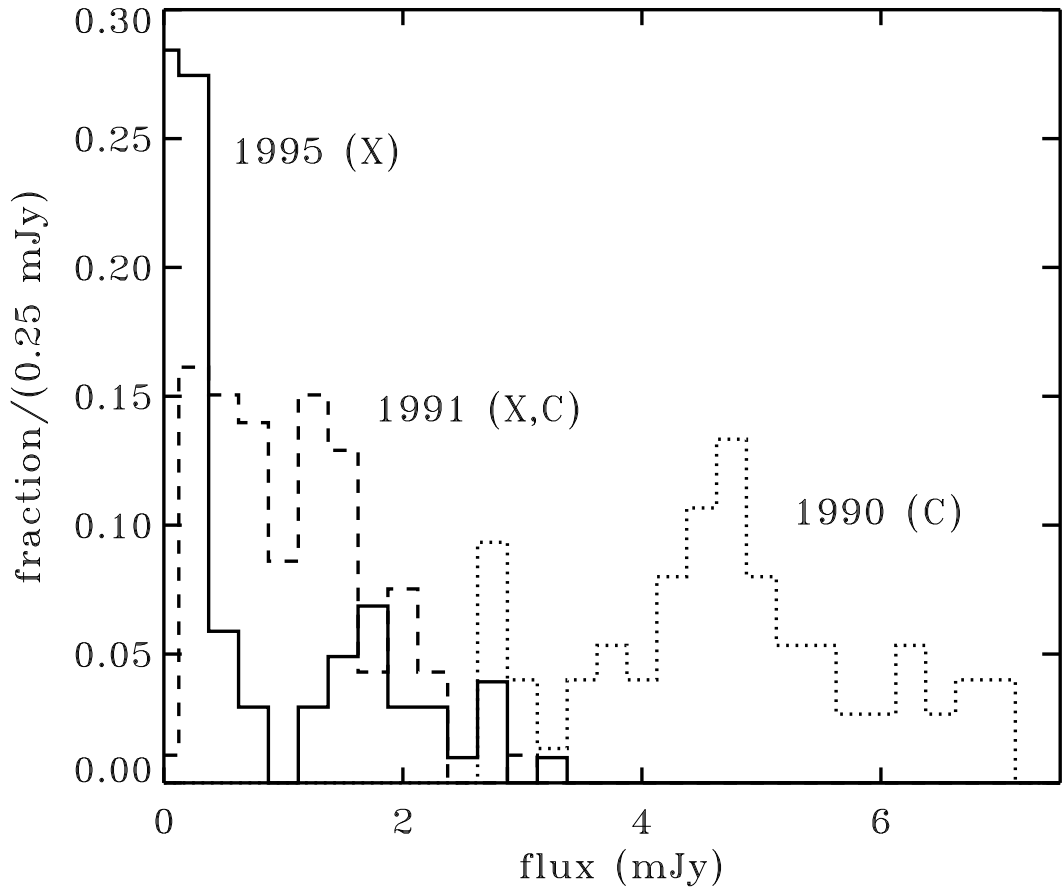,height=4.5in}}
\vskip 0.5in
\caption{\label{fig3}
The histograms of the radio fluxes for observations of ER~Vul
in 1990 (dotted line), 1991 (broken line)
and 1995 (continuous line). Note that the 1990 were obtained
in the C-band (6 cm) whereas the 1991 and 1995 runs were in the
X-band (3.6 cm). Checks conducted during 
the 1991 run indicated that the
C- and X-band flux levels were practically identical at that time.
}
\end{figure}

Concerning the radio flux levels in different years:
It is obviously risky to assume that radio emission observed during 
23 hours-long runs in 1990, 1991 and 1995 represent typical flux
levels. However, if we assume that risk and analyze the observed flux
levels in the form of histograms, as in Figure~\ref{fig3}, we see
a systematic drop over those 6 years. In terms of of the global
averages for each program, the mean flux
levels were 4.86, 1.16, 0.74 mJy, while the median flux levels were
4.80, 1.05, 0.32 mJy. Unfortunately, we have no other data 
on activity of ER~Vul to
correlate these numbers with and to check, if the decrease in the
radio emission was accompanied by similar decreases in spot,
chromospheric or coronal activities in 1995.

\section{CONCLUSIONS}
\label{concl}

Radio observations of ER Vul in 1995 in the continuum at 3.6 cm
gave a very similar picture as during the programs conducted in
1990 and 1991 (R92). Similar type of variability was detected
with the dominant slow, hourly time-scale variations at the flux
levels between non-detection and about 3 mJy. The average
flux levels were the lowest so far observed for the star.
The contemporaneous EUV-continuum (70 -- 140 \AA) observations 
did not show any detectable variability. While the new observations
do not bring any changes to the previous picture established by the
more extensive program R92, they will contribute to future studies
of long-term radio variability in ER~Vul.

\acknowledgments
Special thanks are due to Sandra Scott of the David Dunlap Observatory
for efficient AIPS reductions of the VLA data. Support of the Natural
Sciences and Engineering Council (NSERC) of Canada is acknowledged
here with gratitude.

 
\begin{deluxetable}{crcrcr}     
\tablecaption{3.6 cm VLA Observations of ER~Vul in 1995} 
\tablewidth{0pt}
\tablenum{1}
\tablehead{
  \colhead{JD (IAT)} & \colhead{flux} & \colhead{JD (IAT)} 
& \colhead{flux}  & \colhead{JD (IAT)} & \colhead{flux} \\
  \colhead{--2449980} & \colhead{(mJy)} & \colhead{--2449980}
& \colhead{(mJy)} & \colhead{--2449980} & \colhead{(mJy)}
}
\startdata
  4.404 & $  0.31$~~&   4.757 & $  1.70$~~&   7.572 & $ -0.21$~~ \nl
  4.414 & $  0.25$~~&   4.767 & $  1.57$~~&   7.582 & $  0.06$~~ \nl
  4.424 & $  0.05$~~&   4.778 & $  1.51$~~&   7.594 & $  0.19$~~ \nl
  4.435 & $  0.26$~~&   4.788 & $  1.54$~~&   7.605 & $  0.19$~~ \nl
  4.445 & $  0.30$~~&   4.799 & $  1.48$~~&   7.615 & $  0.22$~~ \nl
  4.456 & $  0.32$~~&   4.810 & $  1.70$~~&   7.625 & $  0.23$~~ \nl
  4.466 & $  0.36$~~&   4.821 & $  1.85$~~&   7.636 & $ -0.19$~~ \nl
  4.476 & $  0.02$~~&   4.831 & $  2.25$~~&   7.646 & $  0.04$~~ \nl
  4.487 & $ -0.13$~~&   4.841 & $  2.30$~~&   7.657 & $  0.27$~~ \nl
  4.497 & $  0.24$~~&   4.852 & $  2.45$~~&   7.667 & $  0.15$~~ \nl
  4.508 & $  0.56$~~&   4.862 & $  2.82$~~&   7.677 & $  0.22$~~ \nl
  4.518 & $  0.41$~~&   4.873 & $  2.71$~~&   7.688 & $  0.16$~~ \nl
  4.528 & $  0.36$~~&   4.883 & $  2.98$~~&   7.698 & $  0.16$~~ \nl
  4.539 & $  0.21$~~&   4.893 & $  3.42$~~&   7.709 & $ -0.03$~~ \nl
  4.549 & $  0.39$~~&   4.904 & $  2.86$~~&   7.719 & $ -0.15$~~ \nl
  4.559 & $  0.35$~~&   4.914 & $  2.80$~~&   7.729 & $  0.16$~~ \nl
  4.570 & $  0.29$~~&   4.924 & $  2.44$~~&   7.740 & $  0.38$~~ \nl
  4.580 & $  0.28$~~&   7.392 & $  0.28$~~&   7.750 & $  0.27$~~ \nl
  4.591 & $  0.20$~~&   7.405 & $  0.04$~~&   7.760 & $  0.05$~~ \nl
  4.601 & $  0.27$~~&   7.416 & $  0.07$~~&   7.771 & $ -0.04$~~ \nl
  4.611 & $  0.48$~~&   7.427 & $  0.00$~~&   7.781 & $ -0.16$~~ \nl
  4.622 & $  0.32$~~&   7.437 & $  0.42$~~&   7.792 & $ -0.01$~~ \nl
  4.632 & $  0.50$~~&   7.447 & $  0.74$~~&   7.802 & $  0.23$~~ \nl
  4.643 & $  0.72$~~&   7.458 & $  0.36$~~&   7.812 & $  0.30$~~ \nl
  4.653 & $  0.84$~~&   7.468 & $  0.03$~~&   7.823 & $  0.26$~~ \nl
  4.664 & $  0.90$~~&   7.479 & $  0.39$~~&   7.833 & $  0.23$~~ \nl
  4.674 & $  0.97$~~&   7.489 & $ -0.02$~~&   7.844 & $  0.35$~~ \nl
  4.684 & $  1.36$~~&   7.499 & $  0.25$~~&   7.854 & $  0.38$~~ \nl
  4.695 & $  1.78$~~&   7.510 & $  0.24$~~&   7.864 & $  0.31$~~ \nl
  4.705 & $  1.84$~~&   7.520 & $  0.46$~~&   7.875 & $  0.67$~~ \nl
  4.715 & $  1.99$~~&   7.530 & $  0.01$~~&   7.885 & $  0.64$~~ \nl
  4.726 & $  1.94$~~&   7.541 & $  0.03$~~&   7.895 & $  1.26$~~ \nl
  4.736 & $  2.04$~~&   7.551 & $  0.24$~~&   7.906 & $  1.78$~~ \nl
  4.746 & $  2.03$~~&   7.562 & $  0.42$~~&   7.916 & $  1.78$~~ \nl
\enddata
\end{deluxetable}

\end{document}